\documentclass{sigplanconf}
\usepackage[utf8]{inputenc}
\usepackage{xspace}
\usepackage{graphicx}
\usepackage{pgfplots}
\usepackage{listings}
\usepackage{url}

\begin{document}
\newcommand{\vectorof}[1]{\mbox{{$#1$} {\sf vector}}\xspace}
\newcommand{\nota}[2]{\mbox{\bf {#1}} {\em {#2}}}
\newcommand{\ff}[0]{\mbox{{\sf FastFlow}}\xspace}
\lstset{language=C++, 
           basicstyle=\ttfamily\footnotesize, 
           keywordstyle=\color{blue}\ttfamily,
           stringstyle=\color{magenta}\ttfamily,
           commentstyle=\color{magenta}\ttfamily,
            numbers=none, 
            numberstyle=\scriptsize,
            stepnumber=1,
            numbersep=6pt,
            showstringspaces=false,
            breaklines=true,    
            frame=lines,
            otherkeywords={[[,]],::}, 
            emph={rpr,kernel,in,out,pipeline,stream,farm,map,sync,async},
            emphstyle={\color{red}\bfseries}
  }

\pgfplotsset{width=\linewidth,compat=newest}

\conferenceinfo{HLPGPU 2015}{Prague, January 19, 2015}
\title{State access patterns in embarrassingly parallel computations}
\authorinfo{Marco Danelutto \& Massimo Torquati}
{Dept. of Computer Science -- Univ. of Pisa}
{\texttt{\{marcod,torquati\}@di.unipi.it}}
\authorinfo{Peter Kilpatrick}
{Dept. of Computer Science -- Queen's Univ. Belfast}
{\texttt{p.kilpatrick@qub.ac.uk}}
\date{October 2015}
\toappear{submitted to HLPGPU 2016}
\maketitle

\begin{abstract}
We introduce a set of state access patterns suitable for managing state in embarrassingly  parallel computations on streams. 
The state access patterns are useful to model typical stream parallel applications. We present a classification of the patterns according to the extent and way in which the state is modified. We  define precisely the state access patterns and discuss possible implementation schemas, performances and possibilities to manage adaptivity (parallelism degree) in the patterns. We present experimental results relative to an implementations on top of the structured parallel programming framework \ff  that demonstrate the feasibility and efficiency of the proposed access patterns. 
\end{abstract}



\section{Introduction} 
Structured parallel programming models have been developed to support the design and implementation of parallel applications. These programming models provide the parallel application programmer with a set of pre-defined, ready to use parallel pattern abstractions that may be directly instantiated, alone or in composition, to model the complete parallel behaviour of the application at hand.
This raises the level of abstraction by ensuring that the application programmer
need not be concerned with architectural and parallelism exploitation issues during
application development. Rather, these issues are dealt efficiently, using the state-of-art techniques, by the framework programmer.
Algorithmic skeletons, first introduced in the early '90s in the field of High Performance Computing \cite{cole-manifesto} led to the development of several structured parallel programming frameworks including Muesli \cite{muesli}, SKEPU \cite{skepu10} and \ff \cite{fastflow-cluj}.
Meanwhile, the software engineering community extended the classic design pattern concept \cite{gamma-book} into the \textit{parallel design pattern} concept  \cite{mattson-book}. Although not directly providing the programmer with ready-to-use programming abstractions
(e.g. via library calls, objects, high order functions) modelling the parallel design patterns, this approach enforced the idea that parallelism may be expressed through composition of well-known, efficient and parametric parallelism exploitation patterns rather than through \textit{ad-hoc} compositions of lower level mechanisms.  The
advantages deriving from structured parallel programming approaches have been clearly identified as a viable solution to the development of efficient parallel application in the well-known Berkeley report
\cite{asanovic}.
 
In the framework of parallel design patterns/algorithmic skeletons, stream parallel computations have been widely employed. Various patterns have been provided as algorithmic skeletons working on data streams, and two of them have been demonstrated particularly useful and efficient, namely the {\it pipeline} and the {\it task farm} patterns. In pipelines, parallel computations are structured as a set of stages transforming input tasks to output results. In task farms, the same ``monolithic'' computation is performed over all the input stream items to produce the output result items \cite{kuchen-farm}.

However, despite the clear utility of such patterns, they have traditionally been studied, designed and implemented as \textit{stateless} patterns, i.e. as patterns where the stages (in a pipeline) or the worker (in farm) processes/thread do not support any kind of internal state nor support accesses to some more generalized notion of ``pattern'' global state. 
This despite the fact there are several well know applications requiring the maintenance of either a ``per pattern'' or a ``per component'' state. 

In this work we focus on task farm computation and discuss {\it stateful} pattern variations of the most general embarrassingly parallel pattern provided by the task farm.
In particular we identify a range of cases from read-only state to  the  case  where  every  computation requires  access  to  the complete global state and in turn updates the global state, which is essentially a sequential pattern. 
We highlight as a key  point the fact that  there  exist  intermediate  cases  where  there are  clearly  defined state  updates  and  yet  parallelism  may  be exploited because of the restricted nature of the update in terms
of state.

The specific contribution of this paper consists  therefore in
\begin{itemize}
\item the introduction of a classification scheme for stateful embarrassingly parallel computations and identification of the conditions under which meaningful speedup may be obtained for each of the classes identified; and
\item experimental results on synthetic cases that  illustrate the utility of our scheme for identifying conditions under which speedup may be obtained.
\end{itemize}

The remainder of the paper is structured as follows: Sec.~\ref{sec:farm} presents the task farm, a pattern modelling  embarrassingly parallel computations over streams. Sec.~\ref{sec:fastflow} briefly discusses \ff, the structured parallel programming framework adopted for our experiments. Sec.~\ref{sec:patterns} introduces the proposed state access pattern classification, and Sec.~\ref{sec:experiments} presents preliminary experimental results achieved using a \ff implementation targeting state-of-the-art multicore architectures. Finally, Sec.~\ref{sec:related} discusses related work and Sec.~\ref{sec:conclu} draws conclusions. 


\section{Embarrassingly parallel computations on stream}
\label{sec:farm}
Embarrassingly parallel computations over streams are defined by providing a function $f$ mapping input data stream items to output data stream items. 
We assume that a stream of data items of type $\alpha$ is to be transformed into a stream of data items of type $\beta$. Thus the function $f$ will have type  $f:\alpha \to \beta$ and the result of the computation over an input stream 
\[ \ldots x_3, x_2, x_1, x_0 \]
will be 
\[ \ldots f(x_3), f(x_2), f(x_1), f(x_0) \]

The ordering of the output items w.r.t. the input ones is not
necessarily preserved.  Input data items are available at different
times: if item $x_i$ is available at time $t_i$, item $x_{i+k}$ will
be available at time $t_i + \Delta, \quad \Delta >0$.

Ideally, if input stream item $x_i$ turns out to be available for computation at time $t_i$, then the output stream item $f(x_i)$ will be delivered to the output stream at time $t_i + t_f$, $t_f$ being the time to compute function $f$. 
Suppose input items appear on the input stream every $t_a$, and assuming use of $n_w$ parallel activities (threads, processes) to compute $f$ over different input stream items, the service time of the embarrassingly parallel computation may be approximated as 
\[ T_s(n_w) = max \{ t_a, \frac{t_f}{n_w} \} \]
and the time spent to compute $m$ input tasks as 
\[ T_c(n_w, m) = m T_s \]

\begin{figure}
\centering
\includegraphics[width=\linewidth]{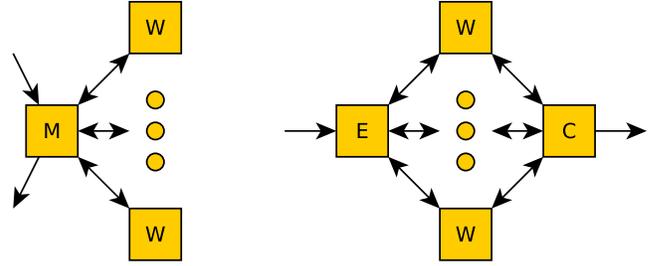}
\caption{Master worker (left) and farm pattern (right)}
\label{fig:farms}
\end{figure}

\paragraph{Implementation}
Embarrassingly parallel computations are usually implemented according to well-know parallel design patterns: 
\begin{itemize}
\item using a \textbf{master/worker} pattern (see Fig.~\ref{fig:farms} left), where a \textit{master} concurrent activity distributes input tasks and collects output results to/from a set of concurrent activities called \textit{workers}. Each worker executes a loop waiting for a task to be computed, computing $f$ and returning the result.
\item using a \textbf{farm pattern} (see Fig.~\ref{fig:farms} right), where an \textit{emitter} concurrent activity schedules input tasks so a set of \textit{workers} computing $f$. Workers in turn direct the output to a \textit{collector} concurrent activity, which in turn delivers the results onto the output stream. In this case the emitter and collector activities are called ``helper'' activities. If the embarrassingly parallel stream computation is not required to enforce input/output ordering of tasks and results (i.e. if $f(x_i)$ may be delivered onto the output stream in any order w.r.t. $f(x_{i-1})$), the collector activity may be suppressed and worker activities may deliver directly to the output stream.
\end{itemize}
In both cases, the master (emitter) concurrent activity may be programmed to implement different scheduling strategies and the master (collector) concurrent activity may be programmed to post-process the $f(x_i)$ items computed by the workers.

\begin{figure}
\begin{lstlisting}
...
#include <ff/farm.hpp>
...
typeOut *worker(typeIn *task, ff_node *const) {
   return ( f(*task) );
}
...
int main(int argc, char * argv[]) {
  ...
  ff_Farm<typeIn, typeOut> myFarm(worker,nw);
  ...
  
  myFarm.run\_and\_wait\_end();
  ...
}
\end{lstlisting}
\caption{\ff task farm code snippet}
\label{fig:ff:farm}
\end{figure}
\section{FastFlow}
\label{sec:fastflow}
In the remainser of the paper, we will discuss possible implementations, while highlighting
advantages and issues, of state access patterns in embarrassingly parallel stream computations on top of \ff \cite{fastflow-web}. 

\ff is a structured parallel programming framework available as an open source, header only library on SourceForge. It has been demonstrated to be very efficient on shared memory multicore machines (possibly with accelerators) and natively provides a number of different stream and data parallel algorithmic skeletons implementing a number of different parallel design patterns \cite{fastflow-cluj}. 

\ff natively provides a \texttt{ff\_farm} class providing the implementation of embarrassingly parallel stream computations, according to the \textit{farm pattern} implementation schema outlined in Sec.~\ref{sec:farm} above. A mandatory emitter thread and an optional collector thread serve as helper threads to schedule tasks to a set of worker threads and to gather, from the workers, results which are eventually dispatched to the \texttt{ff\_farm} output stream. All inter-thread communications are implemented using the \ff lock-free, fast communication mechanisms guaranteeing communication latencies in the order of few (10-40) clock cycles on state-of-the-art multicore systems. 
Emitter scheduling and collector by default implement fair scheduling and gathering policies, while the programmer has the possibility to provide tailored implementations with alternative policies.  
Finally, \ff farms may be equipped with a feedback channel supporting routing back of (partial) results from the collector to the emitter, to implement iterative computations. 

\ff provides the farm pattern to the application programmer as a class that may be directly instantiated to get a parallel application. Fig.~\ref{fig:ff:farm} shows a typical fragment of code which is all that is needed to run a farm pattern in \ff. A farm object may be instantiated providing the input type, the output type, the function computing outputs from inputs and the farm parallelism degree. The execution of the farm is started by calling a \verb1run_and_wait_end()1 method on the farm object. The call is synchronous returning when the farm computation is terminated. Farm objects may be used as components of other patterns. For example the \verb1myFarm1 object in the snippet could have been used as a pipeline stage. In that case the pipeline should have been declared as a pipeline object \verb1myPipe1 and the stage added simply with a \verb1myPipe.addStage(&myFarm);1

The interested reader my find documentation and tutorials on \ff at the \ff web site \cite{fastflow-web}.

\section{State patterns}
\label{sec:patterns}
When state is taken into account in a task farm pattern, different situations may be identified depending on the kind of \textit{state access pattern} used. In the most general and simple case, the computation of the result produced for input item $x_i$ depends on both the value of $x_i$ and on the value of the state at the moment $x_i$ is received. This means that the presence of state serializes the entire computation (see Sec.\ref{sec:serial}).
However, there are several variations of this
computation schema, that turn out to be:
\begin{itemize}
\item useful to model common parallel applications (parallel
  application schemes)
\item supporting a non serial implementation of the state concept or
  at least providing upper/lower bounds on the speedups eventually
  achieved in the computations. 
\end{itemize}
In the following sections, we will first present ``standard'' state pattern in task farm (Sec.~\ref{sec:serial}) as a reference point and then introduce different state patterns of interest in embarrassingly parallel stream computations. Each of the state access patterns will be described by providing precise functional semantics, sample motivating applications, implementation and adaptivity related issues. 

\subsection{Serial state access pattern}
\label{sec:serial}
\paragraph{Motivating example}
A large number of computations require maintenance of a global state to process items from an input stream. For example, transactions issued from different bank operators on the same bank account must be processed such that the bank account details are accessed under mutual exclusive access. 

\paragraph{Definition}
A task farm computation computing the result relative to input task $x_i:\alpha$ as a function of the value of the task and of the value of a state $s:\gamma$ can be formalized by providing two functions $f$ and $s$ such that: 
\begin{itemize}
\item $f: \alpha \time \gamma \to \beta$ computes the result to be delivered to the output stream 
\item $s : \alpha \times \gamma \to \gamma$ computes the new state out of the current task and current state
\end{itemize}
\noindent Then the computation of task $x_i$ requires computing the value to be delivered to the output stream as $f(x_i,s_{i-1})$ and the new state, to be used to compute $x_{i+1}$ as $s(x_i, s_{i-1})$.

Therefore, given an initial state \[s_0:\gamma\] and an input stream \[\ldots, x_2, x_1, x_0\]
the result of the computation  of the task farm may defined as 
\[
    \begin{array}{c}
     \ldots, f(x_2, ns(x_1, ns(x_0, s_0))), 
       f(x_1, ns(x_0,s_0)), f(x_0, s_0) \\
      \end{array}
  \]
  which obviously implies sequential computation of the items
appearing on the output stream.

\paragraph{Implementation}
The serial state access pattern may be implemented using FastFlow by:
\begin{itemize}
\item declaring a global state variable and suitable mutex access mechanisms
\item accessing the global state variable within the worker code (i.e. the code computing $f$) while employing the mutex access mechanisms to guarantee exclusive access to the global state variable. 
\end{itemize}

\paragraph{Performance}
Serial state access pattern, if correctly implemented,  obviously  implies serial execution of the worker code and, as a consequence, any speedup will be achieved using more that a single worker.
%
In Sec.~\ref{sec:separate} we will discuss a slightly different state access pattern which actually provides some possibility for parallelism, while keeping the notion of unique, shared and mutually exclusively accessed global state.

\subsection{Fully partitioned state access pattern}
\label{full:part}
\paragraph{Motivating example}
Deep packet inspection applications need to maintain state relative to each individual connection analyzed. The global state of the deep packet inspection is represented by a vector of states of the single connections. State relative to connection $i$ is only updated when receiving and processing a packet of connection $i$. Incoming packets are processed by different task farm workers, but packets relative to a given connection should be processed by the same worker, the one maintaining the state data structure for that connection.  

\paragraph{Definition}
In the fully partitioned state access pattern the state type is a vector of values of
type $\gamma$ of length $N$ ($vs: \vectorof{\gamma}$) and a function
\[ h : \alpha \to [0,N-1] \]
exists mapping each of the input items to a state vector
position. The state vector is initialized before starting the
computation with some initial value $s_{init} : \gamma$.
Functions $f$ and $s$ are defined as stated in Sec.~\ref{sec:serial} and the computation of the
farm is defined such that 
for each item of the input stream $x_i$, the item output on the
  output stream is \[f(x_i, v[h(x_i)])\] and the state is updated
  such that \[v[h(x_i)] = s(x_i, v[h(x_i)])\]
State items other than $h(x_i)$ are not needed to compute stream item $x_i$.

\paragraph{Implementation}
Given a task farm skeleton with $n_w$ workers, the $N$ state items
will be partitioned among the workers by giving item $v_i$ to worker
$\lceil i/n_w \rceil$. The farm emitter will therefore schedule task $x_k$ to worker
$\lceil h(x_k) / n_w \rceil $ and the worker will be the one hosting the current,
updated value of the state item necessary to compute both the output
result $f(x_i, v[h(x_i)])$ and the state update $v[h(x_i)] = s(x_i,
v[h(x_i)])$.

Finally, the value of the global state may be fetched from the 
farm collector provided the workers direct to the collector their local state items before terminating\footnote{or after a timeout, or after having performed a given number of state updates}. 
Overall, in this implementation, worker $j$ will never be enabled to access state items hosted by the other workers. 


\paragraph{Performance}
Load balancing, and therefore scalability, depends on the efficiency
of the hash function to spread incoming tasks (more or less) equally across the full range of
workers. In the case of a fair implementation of function $h$, close to ideal speedups may be achieved. If the function $h$ directs more items to a subset of the available workers, the speedup achieved will be impaired by a proportional factor. 

\paragraph{Adaptivity}
Increasing the number of workers from $n_w$ to $n_w+1$ requires that worker $i$ directs to worker $w_{i+1}$ 
its last $i+1$ state items: worker $w_0$ directs one state item to $w_1$, $w_1$ directs to $w_2$ 2 items, $w_2$ to $w_3$ 3 items, etc. When decreasing the number of workers from $n_w$ to $n_{w-1}$, worker $w_i$ directs to worker $w_{i-1}$ exactly $i$ state items.

\subsection{Accumulator state access pattern}
\paragraph{Motivating example}
Searching for the number of occurrences of a string in a text (or of DNA sequences in a genome) is a typical application implementing this state access pattern. 

\paragraph{Definition}
In the accumulator state pattern the state is a ‘‘scalar'' value $s:
\gamma$. Functions $f$ and $s$ are defined that compute the result
item and the state update out of the current state and of the current
input item. Function $s$ is restricted to be of the form
\[ s(x_i,s_{i-1}) = g(x_i) \oplus s_{i-1} \] where $\oplus$ is an
associative and commutative operator and $g$ is any function
$g:\alpha \to \gamma$.

\paragraph{Implementation}
A local state value $s_w$ is used by each of the farm workers,
initialized to the identity value w.r.t. function $\oplus$ ($s_{zero}$). The worker
processing item $x_i$ computes $y_i = f(x_i, s_w)$. Then it:
\begin{itemize}
\item either sends $y_i$ immediately to the farm collector, and then
   computes the new state value $s'_w = g(x_i) \oplus s_w$ and
  periodically sends the value $s_w$ to the collector, re-initializing
  $s_w$ to $s_{zero}$; or
\item delivers $y_i$ and $g(x_i)$ to the collector, which will update
  the global state value accordingly, task by task. 
\end{itemize}

\paragraph{Performance}
Load balancing is not affected by the state updates, apart from an
increased load on the collector. Depending on the computational weight
of $\oplus$, the implementation with periodical updates to the
collector will be preferred to the one continuously sending the updates
to the collector.

\paragraph{Adaptivity}
When increasing the number of workers the new workers should be instantiated with a local state value initialized with $s_{zero}$. When decreasing the number of workers, before stopping any worker thread, the locally stored state values should be directed to the collector. 
If workers have to be ``merged'' (e.g. to reduce the worker number but not imposing unexpected update messages on the collector) the resulting worker should be given the ``sum'' of the merged workers local state values ($s_i \oplus s_j$ where workers $i$ and $j$ are merged).

\subsection{Successive approximation state access pattern}

\paragraph{Motivating example}
An application searching a dynamically generated space of solutions for the solution with the best ``fitness'' exemplifies this state access pattern. The global state is represented by the best solution candidate. Both solution and fitness value is stored in the state. Local approximations of the currently available ``best'' solution may be maintained and updated to fasten convergence of the overall computation. Solutions ``worse'' than current ``best'' solution are simply discarded.

\paragraph{Definition}
The pattern manages a state which is a scalar value $s: \gamma$. For an input stream with items $x_i$, a stream of successive approximations of the global state $s_j$ is output by the pattern. Each computation relative to
the task $x_i$ updates state if and only if a given condition
$c:\alpha \times \gamma \to \mbox{\sf bool}$ holds true. In that case
the new state value will be computed as  $s'(x_i, s_{i-1})$. Therefore in this state access pattern we have
\[ s(x_i,s_{i-1}) =  
  \left\{  \begin{array}{ll}
    s_{i-1} & \mbox{\sf iff}\ c(x_i,s_{i-1}) = \mbox{\sf false} \\
    s'(x_i, s_{i-1}) & \mbox{\sf otherwise} \\
  \end{array} \right.
  \]
The state access pattern is defined if and only if $s'$ is monotone in the $s_x$ parameter, that is $s'(x_i,s_{i-1}) \leq s_{i-1}$, and
the computation converges even in the case of inexact state updates, that
is, where different updates read a state value and decide to update
the state with distinct values at the same time (global state updates are anyway
executed in mutual exclusion).

\paragraph{Implementation}
The pattern is implemented with a task farm, where global state value is maintained by the collector. Any update to the state is broadcast to
the workers via a feedback channel to the emitter.  Workers maintain a properly initialized\footnote{e.g. to some known $s_{max}$ value} local copy of the global state $ls:\gamma$. Workers processing an input stream item $x_i$
send update messages ($s(x_i,ls)$) to the collector.
Updates are computed on the local
value of the state, and so this may turn out to be
misaligned with respect to the global state value maintained by the collector and to the local copies maintained by the other workers. 
The collector only accepts state updates satisfying the
monotonic property of $s$, that is if a worker sends an update
which would change the state in a non-monotonic way, that update is
discarded on the basis that a better update has already been found. At any update of its local ‘‘global'' state value, the updated value is output over the pattern output stream, and therefore the pattern output stream hosts all the subsequent successive approximations computed for the global state. 

\paragraph{Performance}
There are three distinct additional overhead sources in the pattern, w.r.t. the plain task farm pattern: 
\begin{itemize}
\item A first performance  penalty is paid to update the global state at the farm collector every time a worker decides to send a state update. As this just requires the comparison among the state currently computed as the ‘‘best'' one in the collector and the update value obtained from the worker, this may be considered negligible.
\item A second performance penalty is paid to send back the global state update to the workers, through the farm feedback channel. This requires an additional communication from collector to emitter and a broadcast communication from emitter to workers. \ff implements both communications very efficiently and so the associated overhead is negligible (in the range of fewer than some hundred clock cycles on state-of-the-art multicore architectures). 
\item A third performance penalty is paid for the extra update messages directed by workers not having available (as local state copy) an updated state value. This happens in the case that the collector has already propagated the new state value but the message has not yet reached the worker. This performance penalty comes in two components: a) the worker may compute an extra $s'(x_i,s_{i-1})$ as a consequence of having a wrong $s_{i-1}$ value  in the computation of $c(x_i,s_{i-1})$, and b) the worker directs an extra state update message to the collector. 
\end{itemize}

\paragraph{Adaptivity}
When the number of workers in the farm is increased, the new worker(s) should be given the current value of the global state maintained in the collector. This can also be implemented by allowing the worker(s) to be started with a proper $s_{init}$  and then leaving the new workers to get regular update values from the collector. This obviously slows down the convergence of the overall computation, as the new workers will initially only provide ``wrong'' approximations of the global state. 
When the number of workers in the farm is decreased, the candidate workers to be removed may simply be stopped immediately before attempting to get a new task on their input stream from the emitter. 

\subsection{Separate task/state function state access pattern}
\label{sec:separate}


\paragraph{Motivating example}
A matrix multiplication implemented by generating a stream of $\langle \mbox{row}_i, \mbox{col}_j \rangle$ reference pairs, applying vector product on each pairs and eventually updating the result matrix (state) in the corresponding $i,j$ position is a representative application of the state access pattern. All isomorphic applications, processing stream of items each contributing to ta global state in a non associative and commutative way are representatives of the pattern as well. 

\paragraph{Definition}
The separate task/state function access pattern implements again a scalar state  $s:\gamma$. The computation relative to the input task $x_i:\alpha$ is performed in two steps: first a function $f: \alpha \to \beta$ (not depending on state values) is applied to the input task to obtain $y_i = f(x_i)$. 
Then, a new global state value $s_i$ is computed out of $y_i$ and of the current value of the global state $s_{i-1}$ 
\[s_i = s(y_i, s_{i-1})\]
The computation of a generic task $x_i$
will therefore require some time ($t_f$) to compute $f$ and \textit{then} some
time to fetch the current state value and to compute and commit the
state update($t_s$). The pattern outputs all modifications applied to the global state $s$ onto the output stream. A variant worth being considered is the one only outputting the value updates to the global state $s_j$ such that $\mbox{\em cond}(s_j)$ holds true for some $c: \gamma \to \mbox{\sf bool}$. 

Overall, this pattern is similar to that discussed in Sec.~\ref{sec:serial} (the ``serial state access pattern''), the main difference being the way in which the global state is accessed to compute input task $x_i$: 
\begin{itemize}
\item in the serial state access pattern, the state is read at the beginning and written at the end of the task computation. 
\item in the separate task/state function access pattern the state is accessed only while computing $s$. For the whole period needed to compute $f$ there is no need to access the global state. 
\end{itemize}

\paragraph{Implementation}
The access pattern is implemented on top of a FastFlow farm. 
A global variable is allocated in shared memory before actually starting the farm, along with all the appropriate synchronization mutexes/locks/semaphores needed to 
ensure mutually exclusive access to the state. Pointers to the shared data and to all the required synchronization mechanism variables are passed to all the parallel components composing the task farm pattern. 
A generic farm worker therefore computes $f$ relative to the received $x_i$ task and then a) accesses shared global state using the synchronization mechanisms provided along with the shared state pointer; b) computes the state update; c) updates the global state; and d) eventually releases the locks over the global state. 

\paragraph{Performance}
Scalability of the separate task/state function state access pattern is obviously impacted by the ratio of the time
spent in a worker to compute $f$ (the $t_f$) to the time spent to
interact with the server to update the state (the $t_{s}$), the
latter contributing to the ``serial fraction'' of the farm.
The time taken to compute $n_w$ tasks sequentially will be $n_w(t_f+t_s)$. The time spent computing the same tasks in parallel, using $n_w$ workers will be (at best) $n_w t_s + t_f$ and therefore the maximum speedup will be limited by
\begin{equation}
 \lim_{n_w \to \infty} speedup(n_w) = \lim_{n_w \to \infty}  \frac{n_w(t_f+t_s)}{n_w t_s + t_f} = \frac{t_f}{t_s}+1 
\label{formula:upperbound}
\end{equation}

\paragraph{Adaptivity}
Increasing or decreasing the number of workers used does not pose any particular issue. Adding a new worker simply requires addition of the worker to the emitter worker queues. Taking away one worker simply requires to stop it while it is waiting for a new task. 

\section{Experiments}
\label{sec:experiments}
We describe several experiments relating to the different state access patterns discussed in Sec.~\ref{sec:patterns} aimed at demonstrating that the patterns actually work and that the performance results are those predicted. 
The first group of experiments have all been performed on an Intel Sandy Bridge architecture with 16 2-way hyperthreading cores on two sockets  running under Linux 2.6.32 using FastFlow version 2.1.0. In the last part of this Section (Sec.~\ref{sec:other:hw}), we will show also results achieved on other architectures, confirming the kind of results which have been achieved on the Sandy Bridge multicore. 
All the experiments have been run using synthetic applications modelled after the state access pattern under examination. Actual computations are dummy computations only, spending time according to the assumed timings for the different functions (e.g. $f$, $s$, $c$, etc.).


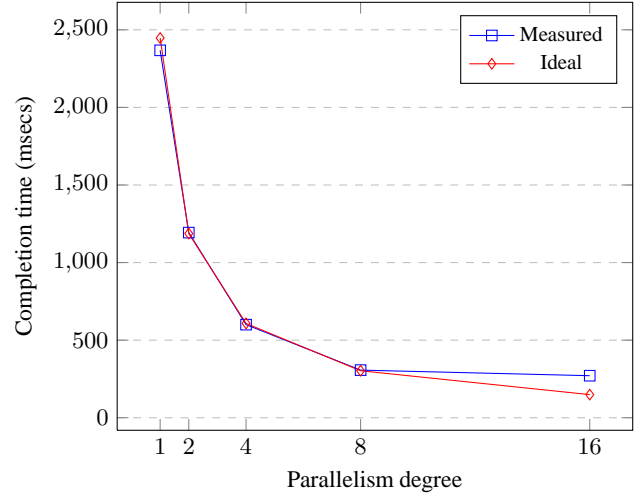
\begin{figure}
\begin{tikzpicture}
\begin{axis}[
    xlabel={Parallelism degree},
    ylabel={Completion time (msecs)},
    xtick={1,2,4,8,16,32,64},
    legend style={legend pos=north east,font=\small},
    ymajorgrids=true,
    grid style=dashed,
]
 
\addplot[
    color=blue,
    mark=square,
    ]
    coordinates {
( 1 , 2367.86 )
( 2 , 1193.18 )
( 4 , 600.055 )
( 8 , 306.945 )
( 16 , 270.922 )
    };
    
\addplot[
    color=red, mark=diamond    ]
    coordinates
    {
( 1 , 2447.36 )
( 2 , 1186.82 )
( 4 , 607.744 )
( 8 , 302.848 )
( 16 , 149.056 )
};
\legend{Measured,Ideal}

\end{axis}
\end{tikzpicture}
\caption{Accumulator state pattern: completion time vs. parallelism degree ($t_f$ 100 times longer than $t_\oplus$)}
\label{fig:small}
\end{figure}

\begin{figure}
\begin{tikzpicture}
\begin{axis}[
    xlabel={Parallelism degree},
    ylabel={Completion time (msec)},
    xtick={1,2,4,8,16,32,64},
    legend style={legend pos=north east,font=\small},
    ymajorgrids=true,
    grid style=dashed,
]
 
\addplot[
    color=blue, mark=square
    ]
    coordinates {
( 1 , 421.397 )
( 2 , 214.329 )
( 4 , 164.671 )
( 8 , 164.11 )
( 16 , 165.169 )
    };
    
\addplot[color=blue, mark=diamond]
coordinates {
( 1 , 421.131 )
( 2 , 213.559 )
( 4 , 109.648 )
( 8 , 84.547 )
( 16 , 85.767 )
};

\addplot[color=blue, mark=otimes]
coordinates {
( 1 , 421.225 )
( 2 , 213.255 )
( 4 , 109.303 )
( 8 , 58.376 )
( 16 , 46.608 )
};
    
\addplot[
    color=red, mark=pentagon
    ]
    coordinates
    {
( 1 , 428.032 )
( 2 , 213.504 )
( 4 , 106.752 )
( 8 , 53.504 )
( 16 , 26.688 )
};
\legend{Measured (freq=1), Measured (freq=2), Measured (freq=4), Ideal}
\end{axis}
\end{tikzpicture}
\caption{Accumulator state pattern: effect of update frequency (considerable state update time ($t_f$ 2 times the $t_\oplus$))}
\label{fig:large}
\end{figure}
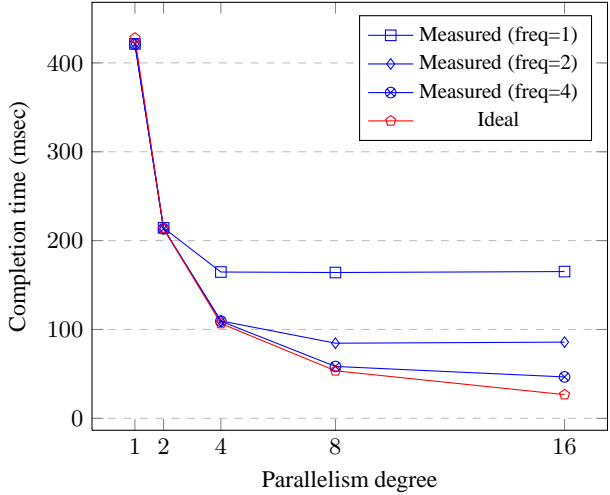

\paragraph{Accumulator state access pattern}
\label{sec:accumulator}
We measured the time spent while running our prototype synthetic application implementing the accumulator state access pattern while varying the amount of time ($t_f$) spent in the computation of the task to be output on the output stream ($f(x_i,s_w)$) and the time ($t_s$) spent in the computation of the new state value/update ($g(x_i) \oplus s_{i-1})$). Fig.~\ref{fig:small} shows the typical result achieved on the Sandy Bridge multicore when  $t_f >> t_s$. In this case, the state access pattern implemented was the one sending regular updates to the collector at each task computation. The $t_f$ was more that 100 times larger than $t_s$ and the completion time for the synthetic application (i.e. the time measured from parallel application start to application end via the \ff function \verb1ffTime1 is almost completely overlapped to the ideal completion time 
\begin{equation}
\frac{m (t_f + t_s) }{n_w} \label{formula:seq}
\end{equation}
Fig.~\ref{fig:large} reports the results achieved when varying the state update message frequency, i.e. when varying the number of task computations awaited (and therefore the number of state updates accumulated to the local state value) before sending the update to the collector, i.e. to the thread maintaining the overall, correct global state of the computation. 
In this case we chose to have $t_f$ close to $t_s$ to stress the effect of collector updates. When sending frequent updates the applications stops scaling at quite small parallelism degrees, while with lower frequency steps scalability comes closer to the ideal. This confirms that, ideally, the frequency update should be chosen to be larger than 
\[\frac{t_f n_w}{t_s}\]
such that when a new update comes to the collector the old ones have been already accumulated in the global state.

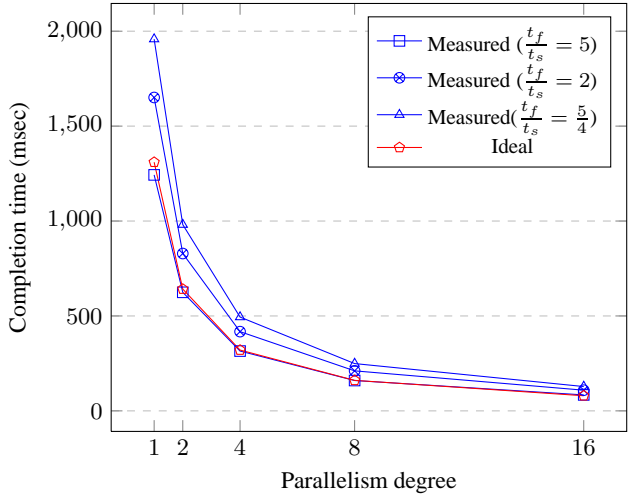
\begin{figure}
\begin{tikzpicture}
\begin{axis}[
    xlabel={Parallelism degree},
    ylabel={Completion time (msec)},
    xtick={1,2,4,8,16,32,64},
    legend style={legend pos=north east,font=\small},
    ymajorgrids=true,
    grid style=dashed,
]
 
\addplot[
    color=blue, mark=square
    ]
    coordinates {
( 1 , 1243.02 )
( 2 , 624.558 )
( 4 , 315.118 )
( 8 , 159.873 )
( 16 , 84.046 )
    };

\addplot[
    color=blue, mark=otimes
    ]
coordinates {
( 1 , 1650.8 )
( 2 , 828.829 )
( 4 , 417.349 )
( 8 , 210.841 )
( 16 , 108.992 )
};
    
\addplot[
    color=blue, mark=triangle
        ]
coordinates {
( 1 , 1958.19 )
( 2 , 980.98 )
( 4 , 494.131 )
( 8 , 249.17 )
( 16 , 128.146 )
};

\addplot[
    color=red, mark=pentagon
    ]
    coordinates
    {
( 1 , 1309.9 )
( 2 , 641.587 )
( 4 , 320.358 )
( 8 , 160.73 )
( 16 , 79.6544 )
};
\legend{Measured ($\frac{t_f}{t_s}= 5$), Measured ($\frac{t_f}{t_s}= 2$), Measured($\frac{t_f}{t_s}= \frac{5}{4}$), Ideal}
\end{axis}
\end{tikzpicture}
\caption{Successive approximation pattern: completion time}
\label{fig:relax1}
\end{figure}

\paragraph{Successive approximation}
Fig.~\ref{fig:relax1} shows results achieved with the implementation of the successive approximation state access pattern on the Sandy bridge architecture. Several curves are plotted against the ideal completion time (the one computing according to (\ref{formula:seq})), varying the amount of time spent computing the condition $c(x_i,s_{i-1})$ ($t_f$ in the legend) and the time spent computing the state update $s'(x_i,s_{i-1})$ ($t_s$ in the legend). As expected, the larger the time spent in the (worker local) computation of the condition, the better the results achieved. 

\begin{figure}
\begin{tikzpicture}
\begin{axis}[
    xlabel={Parallelism degree},
    ylabel={Speedup},
    ymin=0, ymax=400,
    xtick={1,2,4,8,16,32,64},
    ymode = log, 
    ymajorgrids=true,
    grid style=dashed,
    legend style={legend pos=north west,font=\footnotesize}
]
 
\addplot[
    color=blue, mark=square
    ]
    coordinates {
( 1 , 1.00538 )
( 2 , 2.01029 )
( 4 , 4.01494 )
( 8 , 7.98911 )
( 16 , 15.8501 )
    };

\addplot[
    color=blue,domain=1:16   
    ]
{101};

\addplot[
    color=red, mark=square
    ]
    coordinates {
( 1 , 1.00948 )
( 2 , 2.02064 )
( 4 , 4.03186 )
( 8 , 7.98269 )
( 16 , 9.75324 )
};

\addplot[
    color=red,domain=1:16
    ]
{11};

\addplot[
    color=green, mark=square
    ]
    coordinates {
( 1 , 1.00871 )
( 2 , 2.01761 )
( 4 , 4.02362 )
( 8 , 5.7097 )
( 16 , 5.74747 )
};

\addplot[
    color=green,domain=1:16
    ]
{6};

\addplot[color=red, mark=diamond, domain=1:16]{x};

\legend{Measured A, Max Ideal A, Measured B, Max Ideal B, Measured C, Max Ideal C, Ideal}
\end{axis}
\end{tikzpicture}

\caption{Separate task/state function state access pattern: measured vs. ideal speedup}
\label{fig:ser}
\end{figure}
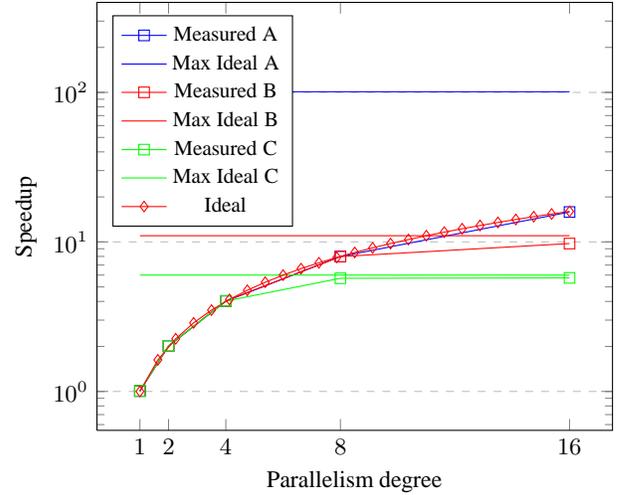

\paragraph{Separate task/state function state access pattern}
The third experiment investigates the performance of the separate task/state function state access pattern. In this case we aimed at verifying if the limits given by (\ref{formula:upperbound}). We therefore run the synthetic application implementing the separate task/state function state access pattern varying the ratio between the time spent computing $f$ and computing $s$. 
Clearly scalability behaves as predicted by (\ref{formula:upperbound}): case A is relative to a situation where the upper bound to the speedup is set to 101 ($t_f = 100 t_s$) and in fact the scalability increases up to the number of available cores as the ideal one. Case B and C are relative to situations where the upper bound is instead 11 and 6, respectively. 

\paragraph{Use of state access patterns in actual applications}
We have no specific scalability/completion time/speedup graphs for non-synthetic applications available at the moment, although we have already some preliminary results achieved with actual application code that will be included in the camera ready of the paper, if accepted. However, this work originated in the activities of our research group and we have already one paper published related to the results achieved with an application \textit{de facto} implementing the partitioned state access pattern \cite{desensi1, desensi2}. This is an application using a partitioned state access pattern implementing a hash function that directs packets to workers respecting the state partitioning schema. The application also supported dynamic adaptation in that the number of workers is increased or decreased to react to packet bursts on the network. Scalability was demonstrated as well as suitability of the dynamic re-distribution of the partitioned state according to the policy outlined in Sec.~\ref{sec:accumulator}.

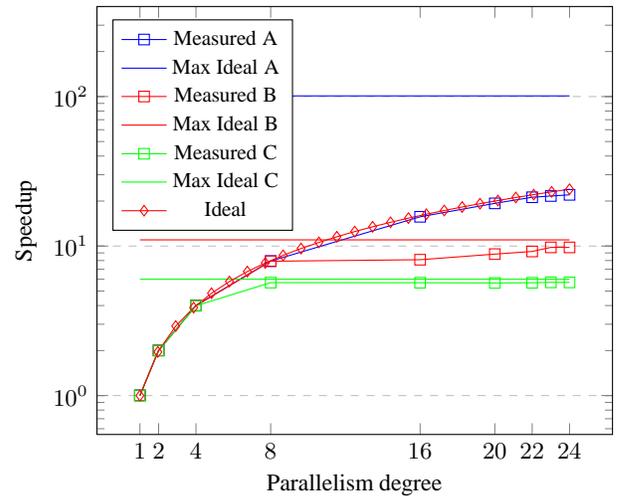
\begin{figure}
\begin{tikzpicture}
\begin{axis}[
    xlabel={Parallelism degree},
    ylabel={Speedup},
    ymin=0, ymax=400,
    xtick={1,2,4,8,16,20,22,24,32,64},
    ymode = log, 
    ymajorgrids=true,
    grid style=dashed,
    legend style={legend pos=north west,font=\footnotesize}
]
 
\addplot[
    color=blue, mark=square
    ]
    coordinates {
( 1 , 1.00452 )
( 2 , 2.00726 )
( 4 , 4.00504 )
( 8 , 7.94206 )
( 16 , 15.6943 )
( 20 , 19.2976 )
( 22 , 21.2007 )
( 23 , 21.6162 )
( 24 , 22.0075 )
    };

\addplot[
    color=blue,domain=1:24   
    ]
{101};

\addplot[
    color=red, mark=square
    ]
    coordinates {
( 1 , 1.00108 )
( 2 , 2.00043 )
( 4 , 3.9928 )
( 8 , 7.91887 )
( 16 , 8.10387 )
( 20 , 8.83359 )
( 22 , 9.20166 )
( 23 , 9.80039 )
( 24 , 9.79522 )
};

\addplot[
    color=red,domain=1:24
    ]
{11};

\addplot[
    color=green, mark=square
    ]
    coordinates {
( 1 , 1.00134 )
( 2 , 2.00029 )
( 4 , 3.98785 )
( 8 , 5.69628 )
( 16 , 5.6858 )
( 20 , 5.668 )
( 22 , 5.6827 )
( 23 , 5.71783 )
( 24 , 5.71341 )
};

\addplot[
    color=green,domain=1:24
    ]
{6};

\addplot[color=red, mark=diamond, domain=1:24]{x};

\legend{Measured A, Max Ideal A, Measured B, Max Ideal B, Measured C, Max Ideal C, Ideal }
\end{axis}
\end{tikzpicture}
\caption{Separate  task/state  function  state  access  pattern:  measured vs. ideal speedup (AMD Magny Cours, 24 cores) }
\label{fig:titanic}
\end{figure}

\begin{figure}
\begin{tikzpicture}
\begin{axis}[
    xlabel={Parallelism degree},
    ylabel={Completion time},
    ymin=0, ymax=18000,
    xtick={1,2,4,8,16,32,64},
    legend style={legend pos=north east,font=\small},
    ymajorgrids=true,
    grid style=dashed,
]
 
\addplot[
    color=blue, mark=square
    ]
    coordinates {
( 1 , 15486 )
( 2 , 7752 )
( 4 , 5196 )
( 8 , 5199 )
( 16 , 5198 )
( 32 , 5213 )
( 64 , 5222)
    };
    
\addplot[color=blue, mark=otimes]
coordinates {
( 1 , 15481 )
( 2 , 7756 )
( 4 , 3907 )
( 8 , 2624 )
( 16 , 2625 )
( 32 , 2630 )
( 64 , 2655 )
};

\addplot[color=blue, mark=otimes]
coordinates {
( 1 , 15479 )
( 2 , 7749)
( 4 , 3893 )
( 8 , 1980 )
( 16 , 1055 )
( 32 , 778 )
( 64 , 800 )
};

\addplot[color=blue, mark=otimes]
coordinates {
( 1 , 15486 )
( 2 , 7749 )
( 4 , 3893 )
( 8 , 1979 )
( 16 , 1054 )
( 32 , 659 )
( 64 , 585 )
};

\addplot[
    color=red, mark=pentagon
    ]
    coordinates
    {
( 1 , 15116 )
( 2 , 7558 )
( 4 , 3779 )    
( 8 , 1889 )
( 16 , 944 )
( 32, 472)
(64, 236)
};
\legend{Measured (freq=1), Measured (freq=2), Measured (freq=4), Measured (freq=8), Measured (freq=16), Ideal}
\end{axis}
\end{tikzpicture}
\caption{Accumulator  state  pattern:  effect  of  update  frequency
(considerable state update time ($t_f$   2 times the $t_\oplus$)) on Power8M, 20 cores, 160 hw contexts.}
\label{fig:power}
\end{figure}
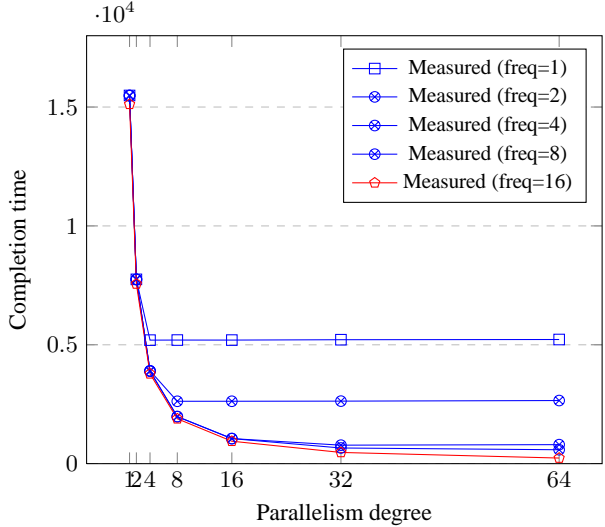

\subsection{Different architectures}
\label{sec:other:hw}
\begin{figure}
\begin{tikzpicture}
\begin{axis}[
    xlabel={Parallelism degree},
    ylabel={Completion time},
    xtick={1,2,4,8,16,32,64},
    legend style={legend pos=north east,font=\small},
    ymajorgrids=true,
    grid style=dashed,
]
 
\addplot[ 
    color=blue, mark=square
    ]
    coordinates {
( 1 , 28790.1 )
( 2 , 15120 )
( 4 , 10299.8 )
( 8 , 10419.8  )
( 16 , 10509.8 )
( 32, 10670.1) 
(64, 10570.1) 
    };
    
\addplot[color=blue, mark=diamond] 
coordinates {
( 1 , 22709.8 )
( 2 , 11950 )
( 4 , 5199.8)
( 8 , 2709.78 )
( 16 , 2879.79)
( 32, 3069.83)
( 64, 2970.09)
};

\addplot[ 
    color=blue, mark=pentagon
    ]
    coordinates
    {
( 1 , 21010.1  )
( 2 , 10560  )
( 4 , 5199.77)
( 8 , 2919.78)
( 16 , 1749.77 )
( 32, 1230.1 )
( 64, 1249.77)
};

\addplot[color=blue, mark=otimes] 
coordinates {
( 1 , 20869.7)
( 2 , 10460.1)
( 4 , 5199.79)
( 8 ,  2719.78)
( 16 , 2059.77)
( 32, 1350.08)
( 64, 796.687)
};

\legend{Measured (freq=1), Measured (freq=4), Measured (freq=16), Measured (freq=32)}
\end{axis}
\end{tikzpicture}
\caption{Accumulator  state  pattern:  effect  of  update  frequency
(considerable state update time ($t_f$   2 times the $t_\oplus$)) on Xeon PHI (60 cores, 4 hw contexts each).}
\label{fig:largephi}
\end{figure}
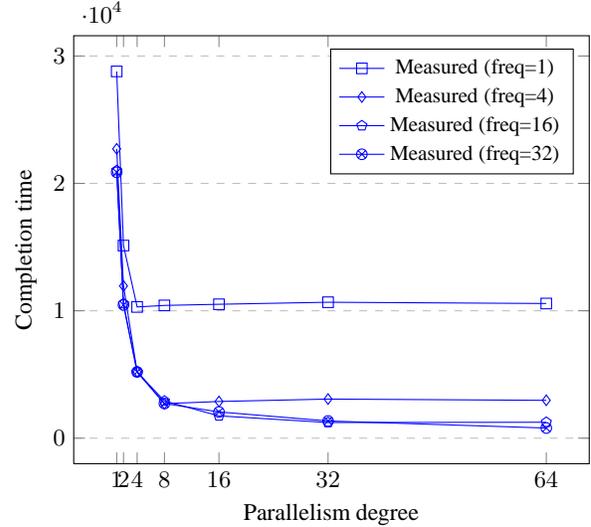

The experimental results discussed so far have all been achieved on the same Intel Sandy Bridge multicore architecture. However, similar results may have been achieved also when running our synthetic applications on different types of state-of-the-art architectures. Plots in Fig.~\ref{fig:largephi}, Fig.~\ref{fig:power} and Fig.~\ref{fig:titanic} show results achieved respectively on a Xeon PHI 5100 architecture (60 cores, 4-way hyper threading), on a IBM Power8E architecture (20 cores, 8 hardware thread contexts per core) and on an AMD Opteron 6176 Magny Cours
 architecture (24 cores).

\section{Related work}
\label{sec:related}

A number of authors have considered various aspects of state in the
context of stream processing. Typically, they employ less overtly structured
approaches than the pattern-based concept presented here. Perhaps the closest to our work is that of Wu et al. \cite{Wu:2012:PSO:2335484.2335515} who introduce a framework for parallelizing stateful operators in
a stream processing system. Their {\it split-(process*)-merge} assembly is
very similar to the task farm presented here. They divide each stateful instance
of {\it process} into non-critical access and critical access segments and
present a more comprehensive theoretical model to determine speedup (based on 
shared lock access times, queue lengths, etc. than is attempted here. However,
they do not attempt the sort of classification scheme given in this work.

Verdu et al \cite{Verdu:2008:MPE:1477942.1477954} focus on implementation issues in relation
to parallel processing of stateful deep pack inspection. The propose Multilayer
Processing as a model to leverage parallelism in stateful applications. They focus
on lower level implementation issues, such as caching and do not explicitly
employ structured pattern based parallelism of the kind used here. 

Gedik \cite{gedik14}
examines properties of partitioning functions for distributing  streaming
data across a number of parallel channels. Thus the author focuses on the equivalent
of properties of the hash function in our fully partitioned state access pattern.

De Matteis et al. \cite{tiziano-popp16} discuss stateful, window based, stream parallel patterns particularly suited to model financial applications. The techniques used to implement the applications fit the design patterns discussed in this paper, but actually somehow mix accumulator, partitioned and separate task/state state access patterns. 

Fernandez et al. \cite{castrofernandez_et_al:OASIcs:2013:4266} also consider the partitioned state and examine issues related
to dynamic scale-out and fault tolerance. As with the others, they do not
use a pattern-based approach nor do they attempt a classification scheme of the kind presented here.

\section{Conclusions}
\label{sec:conclu}
Stream processing has become increasing prevalent as a means to address the 
needs of applications in domains such as network processing, image processing and
social media analysis.
Such applications, when targeted at multicore systems, may be implemented using
task farm and pipeline parallel patterns.
We observe that typically such applications employ task farms in stateless
fashion as it is here that the implementation is easiest and the return in terms
of parallel speedup is greatest.
However, we note that, while embracing state can lead to a de facto sequential computation, there are variations which can provide scope for parallel speedup.
We have classified these variations, indicating for each the issues that arise in relation to implementation detail, performance and how the pattern may be adapted to vary performance.
We have presented experimental evidence that the performance properties the 
various classes of stateful task farms behave as predicted.
We consider that a greater understanding of the
extent to which (streaming) parallel patterns may incorporate state will broaden
the possibilities for development of multicore applications using parallel pattern
based approaches.
To this end, our next step is to investigate other traditionally stateless 
patterns for stateful variants.

\subsection*{Acknowledgments}
This work has been partially supported by EU FP7-ICT-2013-10 project REPARA (No. 609666) ``Reengineering and Enabling Performance And poweR
of Applications'' and EU H2020-ICT-2014-1 project RePhrase (No. 644235) ``REfactoring Parallel Heterogeneous Resource-Aware Applications - a Software Engineering Approach''.

\bibliographystyle{abbrv}
\bibliography{bib}

\begin{thebibliography}{10}

\bibitem{asanovic}
K.~Asanovic, R.~Bodik, J.~Demmel, T.~Keaveny, K.~Keutzer, J.~Kubiatowicz,
  N.~Morgan, D.~Patterson, K.~Sen, J.~Wawrzynek, D.~Wessel, and K.~Yelick.
\newblock A view of the parallel computing landscape.
\newblock {\em Commun. ACM}, 52(10):56--67, Oct. 2009.

\bibitem{cole-manifesto}
M.~Cole.
\newblock Bringing skeletons out of the closet: A pragmatic manifesto for
  skeletal parallel programming.
\newblock {\em Parallel Comput.}, 30(3):389--406, Mar. 2004.

\bibitem{desensi1}
M.~Danelutto, L.~Deri, and D.~D. Sensi.
\newblock Network monitoring on multicores with algorithmic skeletons.
\newblock In K.~D. Bosschere, E.~H. D'Hollander, G.~R. Joubert, D.~A. Padua,
  F.~J. Peters, and M.~Sawyer, editors, {\em Applications, Tools and Techniques
  on the Road to Exascale Computing, Proceedings of the conference ParCo 2011,
  31 August - 3 September 2011, Ghent, Belgium}, volume~22 of {\em Advances in
  Parallel Computing}, pages 519--526. {IOS} Press, 2011.

\bibitem{desensi2}
M.~Danelutto, D.~D. Sensi, and M.~Torquati.
\newblock Energy driven adaptivity in stream parallel computations.
\newblock In M.~Daneshtalab, M.~Aldinucci, V.~Lepp{\"{a}}nen, J.~Lilius, and
  M.~Brorsson, editors, {\em 23rd Euromicro International Conference on
  Parallel, Distributed, and Network-Based Processing, {PDP} 2015, Turku,
  Finland, March 4-6, 2015}, pages 103--110. {IEEE}, 2015.

\bibitem{fastflow-cluj}
M.~Danelutto and M.~Torquati.
\newblock Structured parallel programming with “core” fastflow.
\newblock In V.~Zsók, Z.~Horváth, and L.~Csató, editors, {\em Central
  European Functional Programming School}, volume 8606 of {\em Lecture Notes in
  Computer Science}, pages 29--75. Springer International Publishing, 2015.

\bibitem{skepu10}
J.~Enmyren and C.~W. Kessler.
\newblock Skepu: A multi-backend skeleton programming library for multi-gpu
  systems.
\newblock In {\em Proceedings of the Fourth International Workshop on
  High-level Parallel Programming and Applications}, HLPP '10, pages 5--14, New
  York, NY, USA, 2010. ACM.

\bibitem{muesli}
S.~Ernsting and H.~Kuchen.
\newblock Algorithmic skeletons for multi-core, multi-gpu systems and clusters.
\newblock {\em Int. J. High Perform. Comput. Netw.}, 7(2):129--138, Apr. 2012.

\bibitem{fastflow-web}
{FastFlow home page}, 2015.
\newblock \url{http://calvados.di.unipi.it}.

\bibitem{castrofernandez_et_al:OASIcs:2013:4266}
R.~C. Fernandez, M.~Migliavacca, E.~Kalyvianaki, and P.~Pietzuch.
\newblock {Scalable and Fault-tolerant Stateful Stream Processing}.
\newblock In A.~V. Jones and N.~Ng, editors, {\em 2013 Imperial College
  Computing Student Workshop}, volume~35 of {\em OpenAccess Series in
  Informatics (OASIcs)}, pages 11--18, Dagstuhl, Germany, 2013. Schloss
  Dagstuhl--Leibniz-Zentrum fuer Informatik.

\bibitem{gamma-book}
E.~Gamma, R.~Helm, R.~Johnson, and J.~Vlissides.
\newblock {\em Design Patterns: Elements of Reusable Object-oriented Software}.
\newblock Addison-Wesley Longman Publishing Co., Inc., Boston, MA, USA, 1995.

\bibitem{gedik14}
B.~Gedik.
\newblock Partitioning functions for stateful data parallelism in stream
  processing.
\newblock {\em The VLDB Journal}, 23(4):517--539, 2014.

\bibitem{tiziano-popp16}
T.~D. Matteis and G.~Mencagli.
\newblock {Keep Calm and React with Foresight: Strategies for Low-Latency and
  Energy-Efficient Elastic Data Stream Processing}.
\newblock In {\em {Proceedings of the 21st ACM SIGPLAN Symposium on Principles
  and Practice of Parallel Programming (PPoPP ’16)}}, 2016.
\newblock Barcelona, Spain, to appear.

\bibitem{mattson-book}
T.~Mattson, B.~Sanders, and B.~Massingill.
\newblock {\em Patterns for Parallel Programming}.
\newblock Addison-Wesley Professional, first edition, 2004.

\bibitem{kuchen-farm}
M.~Poldner and H.~Kuchen.
\newblock On implementing the farm skeleton.
\newblock {\em Parallel Processing Letters}, 18(1):117--131, 2008.

\bibitem{Verdu:2008:MPE:1477942.1477954}
J.~Verd\'{u}, M.~Nemirovsky, and M.~Valero.
\newblock Multilayer processing - an execution model for parallel stateful
  packet processing.
\newblock In {\em Proceedings of the 4th ACM/IEEE Symposium on Architectures
  for Networking and Communications Systems}, ANCS '08, pages 79--88, New York,
  NY, USA, 2008. ACM.

\bibitem{Wu:2012:PSO:2335484.2335515}
S.~Wu, V.~Kumar, K.-L. Wu, and B.~C. Ooi.
\newblock Parallelizing stateful operators in a distributed stream processing
  system: How, should you and how much?
\newblock In {\em Proceedings of the 6th ACM International Conference on
  Distributed Event-Based Systems}, DEBS '12, pages 278--289, New York, NY,
  USA, 2012. ACM.

\end{thebibliography}

\end{document}